\documentclass[12pt,preprint]{aastex}

\usepackage{epsf}  
\usepackage{amsmath}  
\shorttitle{$M/L_\star$ ratio of young massive clusters}
\shortauthors{Boily et al.}
\slugcomment{To appear in the ApJ, 10 October issue}

\usepackage{epsf} 
\usepackage{amsmath} 
\usepackage{epic}
\usepackage{graphicx}
\usepackage{epsfig}
\usepackage{rotating}
\usepackage{makeidx}
\usepackage{minitoc}
\usepackage{color}

\begin{document} 
 
% \input{mymacros.tex}  

% MYMACROS.TEX
% uses macros defined in doublespace.sty e.g \singlespace
% Stolen again from MJH Nov 26 1992, subsequent messes are all mine...
\newcommand{\getinptfile}
{\typein[\inptfile]{Input file name}
\input{\inptfile}}
% ================================================
% math mode macros
% force math mode

\newcommand{\mysummary}[2]{\noi {\bf SUMMARY}#1 \\ \noi \sl #2 \\ \capline 
	\hspace{-.13in} \raisebox{.0in}{$\sqcap$} \rm }  
\newcommand{\mycaption}[2]{\caption[#1]{\footnotesize #2}} % First argument:lof
\newcommand{\capline}{\mbox{}\hrulefill}
 % These \my... commands below create \junk.. everytime they're invoked;
 % \junk.. itself is called in mythesis.sty, in the \myheadings definitions;
 % one passes as an argument to \my... the running title of a section, say,
 % which appears in the headings. 
\newcommand{\mysection}[2]{ 
\section{\uppercase{\normalsize{\bf #1}}} \def\junksec{{#2}} } %
\newcommand{\mychapter}[2]{ \chapter{#1} \def\junkchap{{#2}}  % for big headers
\def\thesection{\arabic{chapter}.\arabic{section}}
\def\thesubsection{\thesection.\arabic{subsection}}
\def\thesubsubsection{\thesubsection.\arabic{subsubsection}}
\def\theequation{\arabic{chapter}.\arabic{equation}}
\def\thefigure{\arabic{chapter}.\arabic{figure}}
\def\thetable{\arabic{chapter}.\arabic{table}}
}
\newcommand{\mysubsection}[2]{ \subsection{#1} \def\junksubsec{{#2}} }
	% You can't mark these sections with \footnote; instead 
	% one types \thenote in the (first) bracket, and use \footnotetext
	% outside of the curly brackets to send text to the bottom of the page.
	% the counters are taken care of automatically. 
\def\thenote{\addtocounter{footnote}{1}$^{\scriptstyle{\arabic{footnote}}}$ }

\newcommand{\myfm}[1]{\mbox{$#1$}}
% produce <~ and >~
\def\spose#1{\hbox to 0pt{#1\hss}}	% Definition of .le., .ge. symbols
\def\ltabout{\mathrel{\spose{\lower 3pt\hbox{$\mathchar"218$}} % courtesy AJC1.
     \raise 2.0pt\hbox{$\mathchar"13C$}}}
\def\gtabout{\mathrel{\spose{\lower 3pt\hbox{$\mathchar"218$}}
     \raise 2.0pt\hbox{$\mathchar"13E$}}}
\newcommand{\ltsim}{\raisebox{-0.5ex}{$\;\stackrel{<}{\scriptstyle \backslash}\;$}}
\newcommand{\simlt}{\ltsim}
\newcommand{\simgt}{\gtsim}
%
% units in math mode in roman font
\newcommand{\unit}[1]{\ifmmode \:\mbox{\rm #1}\else \mbox{#1}\fi}
\newcommand{\ze}{\ifmmode \mbox{z=0}\else \mbox{$z=0$ }\fi }
%
% bold vector
\newcommand{\boldv}[1]{\ifmmode \mbox{\boldmath $ #1$} \else 
 \mbox{\boldmath $#1$} \fi}
%
% subscripts with roman font
\renewcommand{\sb}[1]{_{\rm #1}}%
% expectation value
\newcommand{\expec}[1]{\myfm{\left\langle #1 \right\rangle}}
\newcommand{\mone}{\myfm{^{-1}}}
\newcommand{\half}{\myfm{\frac{1}{2}}}
\newcommand{\nth}[1]{\myfm{#1^{\small th}}}
\newcommand{\ten}[1]{\myfm{\times 10^{#1}}}
	% Mathematical definitions / parameters used in the text
\newcommand{\abs}[1]{\mid\!\! #1 \!\!\mid}
\newcommand{\as}{a_{\ast}}
\newcommand{\asr}{(a_{\ast}^{2}-R_{\ast}^{2})}
\newcommand{\bvm}{\bv{m}}
\newcommand{\calf}{{\cal F}}
\newcommand{\calI}{{\cal I}}
\newcommand{\calm}{{v/c}}
\newcommand{\calminf}{{(v/c)_{\infty}}}
\newcommand{\calQ}{{\cal Q}}
\newcommand{\calR}{{\cal R}}
\newcommand{\calw}{{\it W}}
\newcommand{\co}{c_{o}}
\newcommand{\cs}{C_{\sigma}}
\newcommand{\cst}{\tilde{C}_{\sigma}}
\newcommand{\cv}{C_{v}}
\def\dbar{{\mathchar '26\mkern-9mud}}	
\newcommand{\deldelr}{\frac{\partial}{\partial r}}
\newcommand{\deldelR}{\frac{\partial}{\partial R}}
\newcommand{\deldeltheta}{\frac{\partial}{\partial \theta} }
\newcommand{\deldelphi}{\frac{\partial}{\partial \phi} }
\newcommand{\ddotrc}{\ddot{R}_{c}}
\newcommand{\ddotxc}{\ddot{x}_{c}}
\newcommand{\dotrc}{\dot{R}_{c}}
\newcommand{\dotxc}{\dot{x}_{c}}
\newcommand{\Estar}{E_{\ast}}
\newcommand{\grpsi}{\Psi_{\ast}^{\prime}}
\newcommand{\kboltz}{k_{\beta}}
\newcommand{\levi}[1]{\epsilon_{#1}}
\newcommand{\limaso}[1]{$#1 ( a_{\ast}\rightarrow 0)\ $}
\newcommand{\limasinfty}[1]{$#1 ( a_{\ast}\rightarrow \infty)\ $}
\newcommand{\limrinfty}[1]{$#1 ( R\rightarrow \infty,t)\ $}
\newcommand{\limro}[1]{$#1 ( R\rightarrow 0,t)\ $}
\newcommand{\limrso}[1]{$#1 (R_{\ast}\rightarrow 0)\ $}
\newcommand{\limxo}[1]{$#1 ( x\rightarrow 0,t)\ $}
\newcommand{\limxso}[1]{$#1 (\xs\rightarrow 0)\ $}
\newcommand{\ls}{l_{\ast}}
\newcommand{\Ls}{L_{\ast}}
\newcommand{\mean}[1]{<#1>}
\newcommand{\ms}{m_{\ast}}
\newcommand{\Ms}{M_{\ast}}
\newcommand{\nbody}{{\sl N}-body }
\def\nbt{{\sf NBODY2} }
\def\nb1{{\sf NBODY1} }
\newcommand{\nuoned}{\nu\sb{1d}}
\newcommand{\ra}{\rightarrow}
\newcommand{\Ra}{\Rightarrow}
\newcommand{\rc}{r_{c} } % (t)}
\newcommand{\Rc}{R_{c} } % (t)}
\newcommand{\res}[1]{{\rm O}(#1)}
\newcommand{\rnsa}{(r^{2}-a^{2})}
\newcommand{\Rnsa}{(R^{2}-a^{2})}
\newcommand{\rs}{r_{\ast}}
\newcommand{\Rs}{R_{\ast}}
\newcommand{\Rsa}{(R_{\ast}^{2}-a_{\ast}^{2})}
\newcommand{\sa}{\sigma } % [R,t]}
\newcommand{\sac}{\sigma_{c} } % [t]}
\newcommand{\sas}{\sigma_{\ast} } % [R_{\ast}]}
\newcommand{\sasp}{\sigma^{\prime}_{\ast}}
\newcommand{\saxs}{\sigma_{\ast} } % [x_{\ast}]}
\newcommand{\sech}{{\rm sech}}
\newcommand{\tff}{t\sb{ff}} 
\newcommand{\ti}{\tilde}
\newcommand{\trel}{t\sb{rel}}
\newcommand{\ts}{\tilde{\sigma} } % [a,t]}
\newcommand{\tss}{\tilde{\sigma}_{\ast} } % [a_{\ast}]}
\newcommand{\vcol}{v\sb{col}}
\newcommand{\vs}{v_{\ast}  } % [R_{\ast}]}
\newcommand{\vsp}{v^{\prime}_{\ast}}
\newcommand{\vxs}{v_{\ast}  } % [x_{\ast}]}
\newcommand{\xs}{x_{\ast}}
\newcommand{\xc}{x_{c} } % [t]}
\newcommand{\xistar}{\xi_{\ast}}
\newcommand{\rmd}{\ifmmode \:\mbox{{\rm d}}\else \mbox{ d}\fi }
\newcommand{\rmD}{\ifmmode \:\mbox{{\rm D}}\else \mbox{ D}\fi }
\newcommand{\valfven}{v_{{\rm Alfv\acute{e}n}}}

%
% ================================================
%  Abreviations used  in the text ... 
\newcommand{\noi}{\noindent}
\newcommand{\bc}{boundary condition }
\newcommand{\bcs}{boundary conditions }
\newcommand{\Bcs}{Boundary conditions }
\newcommand{\lhs}{left-hand side }
\newcommand{\rhs}{right-hand side }
\newcommand{\wrt}{with respect to }
\newcommand{\iras}{{\sl IRAS }}
\newcommand{\cobe}{{\sl COBE }}
\newcommand{\Oh}{\myfm{\Omega h}}
% \newcommand{\vs}{\vspace{3mm}}
%
% Latin expressions, etc ... 
\newcommand{\etal}{{\em et al.\/ }}
\newcommand{\eg}{{\em e.g.\/ }}
\newcommand{\etc}{{\em etc.\/ }}
\newcommand{\ie}{{\em i.e.\/ }}
\newcommand{\viz}{{\em viz.\/ }}
\newcommand{\cf}{{\em cf.\/ }}
\newcommand{\via}{{\em via\/ }}
\newcommand{\apriori}{{\em a priori\/ }}
\newcommand{\adhoc}{{\em ad hoc\/ }}
\newcommand{\viceversa}{{\em vice versa\/ }}
\newcommand{\versus}{{\em versus\/ }}
\newcommand{\qed}{{\em q.e.d. \/}}
\newcommand{\<}{\thinspace}
%
%%%%%%%%%%%%%%%%%%%%%%%%%%%%%%%%%%%%%%%%%%%%%%%%%%%%%%%%%%%%%%%%%%%%%%
% Astro stuff
%%%%%%%%%%%%%%%%%%%%%%%%%%%%%%%%%%%%%%%%%%%%%%%%%%%%%%%%%%%%%%%%%%%%%%
% units
\newcommand{\km}{\unit{km}}
\newcommand{\kms}{\unit{km~s\mone}}
\newcommand{\kmsa}{\unit{km~s\mone~arcmin}}
\newcommand{\kpc}{\unit{kpc}}
\newcommand{\mpc}{\unit{Mpc}}
\newcommand{\hkpc}{\myfm{h\mone}\kpc}
\newcommand{\hmpc}{\myfm{h\mone}\mpc}
\newcommand{\parsec}{\unit{pc}}
\newcommand{\cm}{\unit{cm}}
\newcommand{\yr}{\unit{yr}}
\newcommand{\au}{\unit{A.U.}}
\newcommand{\AU}{\au}
\newcommand{\gm}{\unit{g}}
\newcommand{\solar}{\myfm{_\odot}}
\newcommand{\solarm}{\unit{M\solar}}
\newcommand{\Lsun}{\unit{L\solar}}
\newcommand{\Rsun}{\unit{R\solar}}
\newcommand{\seconds}{\unit{s}}
\newcommand{\micro}{\myfm{\mu}}
\newcommand{\micrometer}{\micro\mbox{\rm m}}
\newcommand{\Mdot}{\myfm{\dot M}}
\newcommand{\dgr}{\myfm{^\circ} }
\newcommand{\ddgr}{\mbox{\dgr\hskip-0.3em .}}
\newcommand{\mnt}{\mbox{\myfm{'}\hskip-0.3em .}}
\newcommand{\scnd}{\mbox{\myfm{''}\hskip-0.3em .}}
\newcommand{\hr}{\myfm{^{\rm h}}}
\newcommand{\dhr}{\mbox{\hr\hskip-0.3em .}}

	% New environment introduced to cross-reference figures.

\newcommand{\refindent}{\par\noindent\hangindent=0.5in\hangafter=1}
\newcommand{\figpar}{\par\noindent\hangindent=0.7in\hangafter=1}

\newcommand{\mybiblio}{\vspace{1cm}
		       \setcounter{subsection}{0}
		       \addtocounter{section}{1}
		       \def\junksec{References} 
 }

	% References: list of journals.

\newcommand{\vol}[2]{ {\bf#1}, #2}
\newcommand{\jour}[4]{#1. {\it #2\/}, {\bf#3}, #4}

\newcommand{\leftb}{<\!\!} \newcommand{\rightb}{\!\!>} 
\newcommand{\dlam}{\delta\lambda} 
\newcommand{\surl}{\overline}

   \title{The $M/L_\ast$ ratio  of young star clusters in galactic mergers} 
    
   \author{C.~M. Boily, A. Lan\c{c}on} 
\affil{ 
Observatoire astronomique, 11 rue de l'Universit\'e, F-67000 Strasbourg, France; \{cmb,lancon\}@astro.u-strasbg.fr}
% \and 
\author{S. Deiters  and   D.C. Heggie}
\affil{School of Mathematics, University of Edinburgh, King's Building, Edinburgh, Scotland; \{s.deiters,d.c.heggie\}@ed.ac.uk} 
 \received{September 15, 1999; accepted March 16, 2000} 
 
 \begin{abstract}{ 
We point out a strong time-evolution of the mass-to-light conversion
factor $\eta$ commonly used to estimate masses of dense star clusters from
observed cluster radii and stellar velocity dispersions. 
We use a gas-dynamical model coupled with the Cambridge stellar evolution tracks 
to compute line-of-sight velocity dispersions and
half-light radii weighted by the luminosity. 
Stars at birth are assumed to follow the Salpeter mass function in the
range [0.15--17\,M$_{\odot}$].  We find  that $\eta$, and hence the estimated 
cluster mass, increases by factors as large as 3 over time-scales of $20$ million years. 
Increasing the upper mass limit to $50\, M_\sun$ leads to a sharp rise of similar amplitude 
but in as little as $10$ million years. 
 Fitting truncated isothermal (Michie-King) models 
to the projected light profile  leads to  over-estimates of the concentration parameter $c$ 
of $\delta c\approx 0.3$ compared to  the same functional fit applied to the projected mass density. 

}\end{abstract}  
%  
%  14.Sep.'90: Demo-Vs. 
% ________________________________________________________________ 

\section{Introduction}
The formation of star clusters in bursts of star formation during galactic mergers has attracted 
much attention since the ground-breaking study by Schweizer (1986).  
% , for instance through the spatial distribution of interstellar gas or its metal abundance. 
Young clusters can account  for $\sim 20\% $ of the  UV light flux  of their sample starburst 
galaxies (compared with $< 1\%$ for the Milky Way; e.g. Meurer et al. 1995).  
Such numbers bring to focus the role that star  formation in  clusters plays in 
shaping the overall (galactic) stellar mass function. 
%; see e.g. several contributions in Grebel \& Brandner 2002).  
Proto-typical cases where cluster formation has been a spectacular manifestation of interaction-induced starbursts are 
the merging systems NGC 4038/39 (the Antenn\ae) and the nearby galaxy M82 (recent interaction with M81). 
Many of the brightest clusters in these galaxies have estimated ages on the order of $10^7$ years, based on optical and near-IR spectra. 
High resolution spectroscopic data and HST images have been used to measure velocity dispersions and 
estimate virial masses. The line-of-sight velocity dispersion $\sigma_{1d}$  (LOSVD) relates to the 
mass $M$ and projected half-light radius $r_{ph}$ of a cluster in virial equilibrium through 

% \begin{equation} \frac{GM}{r} = \frac{3}{2}\sigma_{1d}^2 \label{eq:Virial} \end{equation}
\begin{equation} 
M = \eta \frac{r_{ph}\sigma_{1d}^2}{G} \ . \label{eq:eta} \end{equation}   
where $\eta$ is a dimensionless free parameter. 
% Spitzer (1987) or Harris (1991) ??? gives $\eta \approx 8$ to 10 for 
% isothermal models of a wide range of concentration $c \equiv \log | r_t/r_o|$, 
% where $r_t$ is the truncation 
% radius and $r_o$ the core or King radius, within which the mass density is nearly constant.  
A number of authors have set $\eta \approx 10 $ 
in their studies  to derive  $M$ from (\ref{eq:eta}) (Sternberg 1998; Mengel et al. 2002; Smith \& Gallagher 2001; McCrady et al. 2003; Maraston et al. 2004). McCrady et al give a derivation
of this value of $\eta$.  Mengel et al. (2002) quote a range from $5.6 - 9.7$ for
King (1966) models with concentration parameter in the range 0.5 to 2.5,
which corresponds to most galactic globular clusters.  We discuss the value of $\eta$ further below. 
%  though acknowledging that variations of $\eta$ immediately  would affect the mass estimate. 
Dynamical masses derived from (\ref{eq:eta}) have been compared to the stellar masses of synthetic 
populations, using both standard (field) and non-standard stellar initial mass functions (henceforth IMF; see Kroupa 2002  for a review). The data were found to be inconsistent with a universal IMF (Mengel et al. 2002; Smith \& Gallagher 2001). In particular, several clusters in M82 were found to be over-luminous with respect to their estimated 
mass (low mass-to-light ratio, $M/L_\ast$). This suggests that M82 clusters may form with a top-heavy stellar IMF (Smith \& Gallagher 2001; McCrady et al. 2003).  

These conclusions hinge on the precise value of $\eta$.  The above studies have assumed no
time-variation of $\eta$, based on the belief that the structural
parameters of young clusters are unlikely to have changed since the
time they were born.  A further implicit assumption made when applying
 (\ref{eq:eta}) is that the stellar subpopulations sampled are all equally representative of the dynamics as a whole. 
This simplification is normally justified on the grounds that the estimated ages of the clusters are too short to allow for 
gradients in their spatial distribution and kinematics to develop from internal evolution. 
 However this train of thought stems from the derivation of a long  relaxation time $t_{rh}$  
for  single-population clusters. Expressed in terms of  the 
dynamical time $t_{cr}$ evaluated at the half-mass radius $R_h$ (Meylan \& Heggie 1997, \S7)

\begin{equation} \frac{t_{rh}}{t_{cr}} \simeq 0.138 \left[ \frac{R_h}{R_g}\right]^{3/2} 
\frac{N}{\ln 0.4 N} \label{eq:trhtrc} \end{equation} 
 where $R_g = GM^2/|W|$ is the gravitational radius and  the ratio $R_h/R_g \approx$ unity for a wide number of model fits to observed  clusters.  With a spectrum of masses, the trend toward equipartition of kinetic energy speeds up evolution, and mass segregation 
now develops on a time-scale given by (Farouki \& Salpeter 1982; Spitzer 1987) 

\begin{equation} \frac{t_{ms}}{t_{rh}} \approx \frac{\pi}{3} \frac{\langle{m}\rangle}{{\rm max}\{m\}} \, \frac{\overline{\rho}}{\rho}\left[ \frac{R_h}{R_g}\right]^{3/2}\, 
\label{eq:two-body} \end{equation} 
where $\overline{\rho}\equiv M/2/(4\pi R_h^3/3)$ is  the  mean density inside the half-mass radius  (an  over-line denotes averaging over space, and  brackets averaging by mass).  
% In practice star clusters span a wide range in local density (Djorgovski 1993) so  (\ref{eq:two-body}) should be applied with discretion. 
% Equation (\ref{eq:two-body})   shows that the local mass-segregation and global relaxation times 
% will differ by large factors whenever the stellar mass function is well-sampled. 
% \footnote{Note that we have presumed equipartition of kinetic energy when 
% deriving (\ref{eq:two-body}); the limit when all populations have identical mean velocity 
% dispersion only introduces  a factor $2\sqrt{2}$ 
% to the numerator.}.  
%  At the half-mass radius, $t_{rh}$ 
%  takes  values of several hundred million years for globular clusters of median membership 
%  $N \simeq 300,000$. 
 %  For this reason two-body relaxation and mass segregation 
 % effects are typically dismissed 
 % in the context of massive cluster formation when the age of the clusters falls below 100 Myrs. 
The numerical coefficients entering (\ref{eq:two-body}) are derived from a stellar IMF  and mass distribution. 
The Galactic-field IMF covers a range from $\approx 0.08\, \solarm$ 
to $\gtabout 60\, \solarm$ (O-stars) and possibly all the way to $\approx 100\, \solarm$ (Kroupa 2002). The mean mass  for this IMF is $\langle{m}\rangle \approx 1.33 \solarm$   
 and therefore the ratio $\langle{m}\rangle/{\rm max}\{m\} $ lies in the range  0.013 -- 0.022 for the chosen upper limits. This dramatically 
reduces the mass-segregation time-scale (\ref{eq:two-body}) and bears on the 
% The consequences this has on the 
parameter $\eta$ (through time-variation of $r_{ph} $ and $\sigma_{1d}$) % need to be 
since the brightest, most massive member stars are also those that undergo the 
most significant segregation and inward migration. 
% This would both bias the projected half-light radius $r_{ph}$ 
% and mean LOSVD to lower values. 
The purpose of this Letter is to show with numerical modelling that, 
  through this process, the mass conversion factor $\eta$  of massive clusters evolves by a  factor 
  of a few over periods as short as 20 millions years, contrary to  expectations of 
  zero evolution. 
 
% We discuss details of  the  numerical 
% method before presenting our results. Applications and future work are discussed in the closing 
% section. 

\section{Numerical method : GasTel} \label{sec:Method}
\noindent The equations of motion were integrated numerically based on  the gas-dynamical approach pioneered by Larson (1970) and developed further by Louis \& Spurzem (1991) to include anisotropic velocity fields.  The method 
leans on an analogy between exchange of kinetic energy through star-star interactions and the classical heat-diffusion process of fluid dynamics (Lynden-Bell \& Eggleton 1980; Heggie \& Ramamani 1989). 
The implementation in spherical coordinates we used is largely due to Louis \& Spurzem (1991) and Giersz \& Spurzem (1994) but extended to include a spectrum of stellar masses  (Spurzem \& Takahashi 1995). % Important updates are discussed in the next paragraph. 
% The code GasTel makes  no attempt  to quantify the effects of external tidal fields. 

The mass spectrum is sampled at constant logarithmic increments in the interval $\{m_0,m_1\}$; we used 
14 bins in our standard runs ($\delta\ln m \approx 0.329$). 
 % To each mass bin corresponds the same initial  (continuous) radial density profile.  
 Star-star
interactions (including those between stars of the same mass bin) lead to  the diffusion of kinetic energy. 
Roughly speaking, the resulting change in the velocity dispersion of each mass
bin causes a readjustment of the density profile.  This is obtained using a semi-implicit Henyey integration. The 
gravitational potential is then updated by applying Poisson's equation (see Louis \& Spurzem 1991). 

We checked that the  mass-segregation time (\ref{eq:two-body}) obtained for GasTel  is in quantitative agreement with N-body and Fokker-Planck integrators   by 
computing the evolution of Plummer models with three species of stars.  
% and following the evolution of Lagrangian radii in time for all three components 
(cf. Spitzer \& Shull 1975, Fig.~1). Spurzem \& Takahashi (1995) report excellent agreement with 
(3) from their two-component test calculations. 

\subsection{Stellar evolution}  
The different mass components are evolved according to the Cambridge tracks (Pols et al. 1998; Hurley et al. 2000).
The  tracks are efficiently coded in the form of fitting functions of the kind first presented 
by Eggleton, Fitchett \& Tout (1989). The functions return the current bolometric luminosity, radius, mass and 
metal abundance for given time and initial metal abundance $z_o$; 
in this contribution we set $z_o = 0.02$ (solar abundance) throughout. 

A filter can be applied to the bolometric luminosity from model stellar atmospheres, and 
% For  given initial  mass and evolutionary epoch, 
the total flux in a specified waveband read % by interpolating polynomials 
 from the Basel stellar library (Lejeune et al. 1998). We have 
 mapped the stellar luminosity near the strongest near-IR CO bandhead 
 ($\lambda \simeq 2.2 - 2.29 \mu m$) and the CaII triplet ( $\lambda \simeq 8200 - 8600 \AA$). 
% together with  the bolometric limit (all wavelengths). 
% We verified that the bolometric limit leaves the Cambridge luminosities unchanged. 

\section{Flux-weighted scheme \& stellar IMF} \label{sec:Flux}Low 
gradients in metal abundances in young clusters can be interpreted to imply a spatial 
distribution of stars independent of their mass at the time the clusters formed  
 (e.g. Suntzeff  1993). All the calculations presented here have no built-in 
segregation initially. 
 
The flux-weighted LOSVD is obtained, first  by averaging the square velocity  dispersion within a projected  radius $R$ for a given stellar mass~;  then by averaging over all stellar masses, using  as statistical weight the total  luminosity $\Lambda_\lambda^{\dlam}(m)$ of mass component $m$ 
over a wavelength  interval $\dlam$ centered on  $\lambda$.  The mean square velocity of stars of mass $m$, $ \surl{\sigma_{m}^2} $,  is computed from their surface density $\Sigma_m$ 

\[
 \surl{\sigma_{m}^2}  (R) = \left.{\displaystyle{\int_0^{R}\sigma_m^2(r)  \Sigma_m(r)  \pi d r^2}}\right/{\displaystyle{\int_0^{R} \Sigma_m(r)  \pi d r^2}} \]
The dispersions  $\surl{\sigma_m^2}$  are summed  over all masses using the luminosity as statistical weight, 

\begin{equation} 
 \langle\sigma_{\lambda}^2\rangle  =\left.\int_{m_0}^{m_1}  \surl{\sigma_{m}^2}
 \Lambda_\lambda^{\dlam}(m[t]) n(m) {\rm d} \ln\, m \right/  {\cal W}_\lambda\, \label{eq:meanflux} 
\end{equation} 
where $n(m)$  is the number of stars in the logarithmic mass interval $\ln\, m, \ln\, m + \delta\ln m$. 
The  normalisation constant ${\cal W}_\lambda$ is the total light flux at wavelength $\lambda$ 
from all stars at time $t$. Note that $n(m)$ in (\ref{eq:meanflux}) refers to the IMF. All time-dependencies  are encapsulated in $\Lambda_\lambda^{\dlam}(m[t])$ and $ \surl{\sigma_{m}^2}$. 
%
%\begin{equation}
% {\cal W}_\lambda \equiv \int_{m_0}^{m_1}  \Lambda_\lambda^{\delta\lambda}(m[t])  n(m) d\ln\,  m \label{eq:Weight}
% \end{equation} 
% where we noted the implicit dependence on time of the stellar masses $m$ ; from (\ref{eq:Weight}) 
% the system LOSVD in (\ref{eq:meanflux}) follows. 

% The projected half-light radius $r_{ph}$ is found by bisection  through a radial  integral similar
%  to the radially averaged dispersion but replacing 
% $\sigma^2_m$ with $ \Lambda^{\delta\lambda}_\lambda $  and summing over all stellar masses in the interval  $\{m_0, m_1\}$. 
%
% \begin{equation} 
% \half = \sum_m \frac{\displaystyle{\int_0^{r_{ph}}\Lambda_\lambda^{\delta\lambda}(r)  
% \Sigma_m(r)  \pi d r^2}}{\displaystyle{\int_0^{\infty} \Lambda_\lambda^{\delta\lambda}(r) \Sigma_m(r)  
% \pi d r^2}} \label{eq:rph} \  ,
% \end{equation} 

% \subsection{Stellar initial mass function}  
% The distribution of stellar masses $n(m)$ is drawn from the {\it initial} mass function.  
For the mass range $1-60\, M_\odot$ (up to O-stars), the single-star IMF is well represented by a   
 Salpeter (1955) power-law, $ n(m) {\rm d} \ln\, m \propto  m^{-2.35} {\rm d} m $. 
 We set the mass range  $\{m_0,m_1\}$  to cover  
two decades, from 0.15 to 17 $\solarm$  (spectral type M5 to B5).  
Note that solar neighbourhood data favour an IMF that flattens out below $1\,\solarm$ 
(Scalo 1986; Kroupa et al. 1993), so our choice of an Salpeter IMF may seem artificial.  
However for the chosen mass range, we compute $\langle{m}\rangle / m_1 \simeq 0.025$, a ratio nearly identical  to that  obtained from the Galactic-field IMF.   
Thus our choice of mass function and mass range yields a conservative mass-segregation 
timescale compared with other IMF's when the latter are extended beyond  $60\, \solarm$. 
% In this sense we {\it de facto} limit the time-evolution of the mass to light ratio $\eta$, so  favouring  the common practice of adopting a constant value. 
 
%   This point will be important later when assessing the results of our investigation. 

\section{Results}\label{sec:Results} 
The runtime of our calculations was set to 50 Myrs. 
All  models were scaled to $N = 500,000$ member stars and a total
 mass of $M \simeq 2\, 10^5 M_\odot$.  
% This is  significantly less than what is  measured  for massive 
% clusters in starburst galaxies, where membership may well exceed $10^6$. 
% This point is important because   the time-scale for segregation scales with $N/\ln N$ (cf. Eq. [2] and [3]). 
%   However we may explore the importance of varying  the  dynamical evolution time by constructing models with 
  We vary the  dynamical evolution time between runs by constructing models with 
  different sizes and central densities, while keeping the total mass, and $N$, constant. 
% This approach is 
%  attractive because two systems with the same mean density will have the same dynamical time $t_{cr}$,  independently of their mass and size.  
   Mengel et al. (2002) fit  the  light profile of young Antenn\ae\ clusters with King models of dimension-less parameter $\Psi/\sigma^2 = 6 $ to $9$.   We first setup  three  models with identical $\Psi/\sigma^2 = 6$  King parameter but each with different half-light radius (Table 1). 
  The model labeled `M82-F' has a mean surface density $\approx 1.5\, 10^4 M_\odot/pc^2$ 
   or half  the value we derive for  that cluster from the data of Smith \& Gallagher (2001). The densest model is labeled `R136', in reference to the 30 Doradus cluster  (central volume density $\sim 10^6 M_\odot/pc^3$;  Brandl et al. 1996).  
  Finally, a third low-density model was evolved for comparison (labeled `Low').  
    
% However doubling 
% $N$ but keeping constant the radius of the cluster would lead to an increase of $t_{rh}$ in (2) by no more than 
% a factor $\approx\sqrt{2}$. A crude scaling up of our calculations to larger systems would be to 
% shift the time axis to larger values by that same factor. 

\begin{figure*}[t]
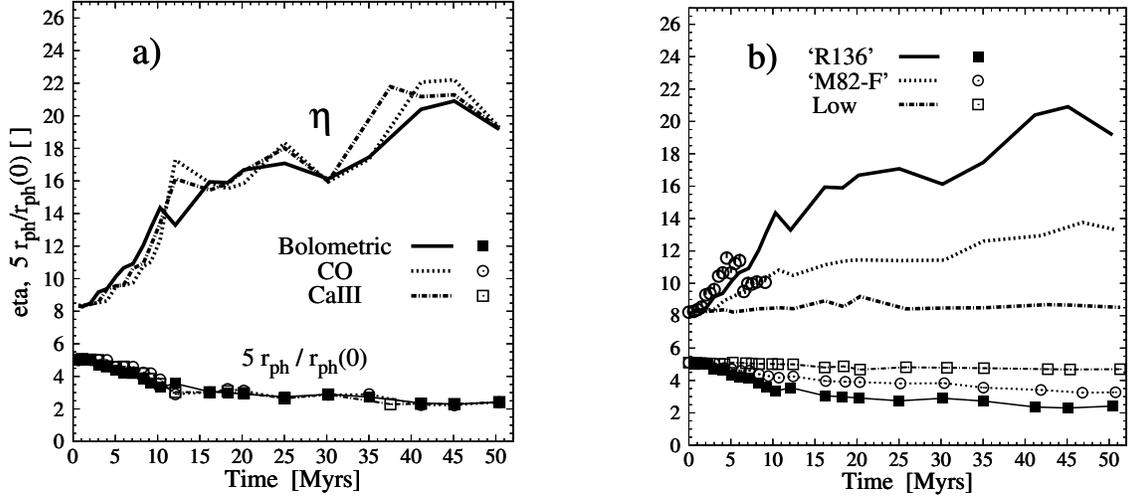

\begin{center} \begin{picture}(100,250) 
\put(-180,0){\epsfig{file=BoilyFig1.epsi,width=0.52\textwidth}}
%%  \put(0,140){\huge $\eta$} \put(-30,60){\Large $5 r_{ph}/r_{ph}(0) $ } 
\put(60,0){\epsfig{file=BoilyFig1b.epsi,  width=0.5\textwidth}}
\end{picture}
\caption{Parameter $\eta$ and half-light radius $r_{ph}$ versus time. The initial models were 
unsegregated King $\Psi/\sigma^2 = 6$ models. a) results 
 using filters at three different wavebands (see text for details); b) results for models with different initial surface density (cf. Table 1). When the mass range is increased from 17 $M_\sun$ to $50 M_\sun$ 
  $\eta$ rises sharply over the first 10 Myrs (open circles, case `M82-F'); thereafter it rejoins  
 the curve displayed. } 
\label{fig:etavstime} \end{center} \end{figure*}

% \subsection{Time-evolution of $\eta$} 
% \subsubsection{Observing `R136' at different wavebands} 
We start with the densest model `R136' for which we compute the shortest relaxation time 
$t_{rh}$. 
Since $M, r_{ph} $ and $\langle\sigma^2_\lambda\rangle$ are all known from the simulations,  we solve for $\eta$ directly from (\ref{eq:eta}). 
The run of $\eta$  in time is displayed on Fig.~\ref{fig:etavstime}(a). 
Three  curves are shown, corresponding to  bolometric light and  filters centered on the 
CO continuum and the CaII triplet. The rapid increase of $\eta$ from an initial value $\approx 8.2$ is striking. 
After 15 Myrs of evolution, $\eta$  has more than doubled.  The projected half-light radius $r_{ph}$ is displayed 
alongside $\eta$. In all the cases $r_{ph}$ decreases steadily in time, a direct result of the 
 migration of massive stars toward the centre. The general 
trend and quantities are not sensitive to the waveband adopted, for both $\eta$ and $r_{ph}$.  Note the  slow but systematic rise of $\eta$ after $\approx 35$ Myrs, when $\eta > 20$  (factor $>$ 2.5 from  its initial value; Fig.~1[a]).  

The LOSVD changes relatively little over time in comparison: we measure a monotonic 
decrease of $\langle\sigma_{\lambda}\rangle$ from  $\approx 10.4$ to 9.3 km/s (or, -10.6\%) for the system as a whole, although 
 for individual components evolution was more significant: down $\approx 30\%$ for 
the most massive stars, while the lightest component enjoys an increase of a comparable 
magnitude.  These effects can all be traced back to the dynamical mass segregation. 

% The ratio of LOSVD's between 
% stars of different mass reached $\approx 1.8$ at most  over the full  
% time of the computation. 

% \subsubsection{The cases of `M82-F' and `Low'} 
% The longer time-scales for segregation of the less dense models `M82-F' and `Low' 
% lead to reduced evolution over 50 Myrs. 
The time-evolution of $\eta$ for all three models is  displayed on Fig.~1(b); we have plotted only the results for 
the bolometric light for clarity.  Both $\eta$ and $r_{ph}$ for the `Low' model 
stay essentially constant throughout. However the case `M82-F'  (initial surface density $> 10 M_\sun$/pc$^2$) shows  unmistakable  evolution,  suggesting that for clusters of such or higher 
 mean surface density we may no longer presume 
a time-independent mass-to-light parameter $\eta$.  

\section{Discussion and future work}
When the projected density of massive star clusters exceeds
 a few $\times 10^4 M_\odot/pc^2$ , the mass-to-light conversion factor $\eta$ used to derive the  
mass in equation (\ref{eq:eta}) may increase by a factor as large as 3 over time-scales of a few  
$\times 10^7$ years (cf. Fig. 1[b] and Table~1).  Clusters with low surface density 
 have long mass-segregation time-scales and constant $\eta$.  When we 
 compare clusters with similar mean surface density $\approx 4\, 10^4 M_\odot/pc^2$  but different central density initially, 
 the results indicate that the  increase in $\eta$ is an increasing function of the  initial central density (Table 1).  Thus the drift of heavy 
 stars toward the centre can considerably bias mass estimates of  clusters centrally peaked at birth. 
   The resolved cluster R136 (NGC 2070)  in the 30 Doradus complex is a case in point: its central volume density may be as high as 
  $\sim 10^6 M_\odot/pc^3$ with an  estimated spherical half-mass radius $\sim 1$ pc (Brandl et al. 1996). % Its  total mass  $\sim 2\, 10^4 M_\odot$ implies a 
 %  projected surface density $ < 10^4 M_\odot/pc^2$ implies a short relaxation time derived from (2).  
 Several clusters in the Antenn\ae\ or M82  have estimated masses $\sim 10^6 M_\odot$ and a projected half-light radius $r_{ph} \sim $ a few pc's.  Smith \& Gallagher (2001) derive $M = 1.2\ 10^6 M_\odot$ and $r_{ph} \approx 2.8$ pc for the cluster M82-F. 
The  mean surface density of this cluster $\sim  2.4\, 10^4 M_\odot/pc^2$ 
 inside $r_{ph}$, comparable to our model `M82-F' ; for comparison, 
 our model `R136' had a mean density twice as high.  All calculations  were done for 
 a membership of $N = 500,000$, a factor 2 lower than rich starburst  clusters. 
 This may serve to accelerate evolution compared with actual massive clusters (cf. Eq. [2]). 
 We recall that our choice of  an  upper mass $m_1 < 20\, \solarm$ 
   yields a mass segregation timescale which is a factor of 3 longer than if we had included 
   O-stars. A re-run of the case `M82-F' with a  mass spectrum widened to 50 $M_\sun$ 
   yielded a rapid increase of $\eta$ over a shorter, 10 Myrs-timescale  (open dots, Fig. 1[b]). 
   Therefore the evolution of $\eta$ through mass segregation of stars  is a sensitive function of both the IMF and 
   the cluster surface density profiles.  We mention that the cluster  M82-F has an age of $\simeq 60 \pm 20 $ Myrs, by  which time the increase of $\eta$ is maximum. 
  This larger value of $\eta$ would go some way toward solving the apparent 
over-luminosity of young M82 clusters with respect to a standard IMF. Recent work 
on M82-F has highlighted the possibility of strong mass segregation in that cluster 
(McCrady et al. 2004). 

The strong evolution of $\eta$  for model `R136' (Fig.~1[a]) suggested to us to compare 
functional fits to  the total surface density and  luminosity profiles. 
 We performed  least-square fits (Press et al. 1992) using King profiles 
 on the simulation grid out to $5 r_{ph}$. 
Doing this for model `R136' after 35 Myrs of evolution we found a best fit to the 
radial luminosity profile  (cf. \S3) that gave a   King parameter $\Psi/\sigma^2 \approx 7.6$ 
(compared with $\simeq 6$ for the run of density).  
The concentration parameter $c = \log|r_t/r_o|$  increases  by $\approx 0.45$ as $\Psi/\sigma^2$ runs  from 6 to 7.6. Since the truncation radius $r_t$ remains constant, 
the central region  seemingly shrinks by a factor $\approx 2.8$ compared with the surface density.  
Table 1  gives values of $\langle\eta\rangle$ and $\langle r_{ph}\rangle$ averaged over the last 25 Myrs of evolution for this and two other models, which also show increased concentration.  
The effect can be cast in the context of  a whole population of clusters as 
detected in interacting galaxies such as  the Antenn\ae\ (NGC 4038/9) or M82. 
Our modelling indicates that
the clusters  should be yet more massive than estimated until now, and less concentrated. 
Therefore their binding energy per unit mass will be less than deduced from profiling the  light. 
   However it has not escaped our attention that fitting the light flux 
would require a convolution with  the point-spread function of an instrument to allow for direct comparisons  with observations. More detailed models will be presented in a follow-up study 
(Fleck et al., in preparation). %  and derive quantities applicable to their future evolution.

\section*{Acknowledgments}  We thank the anonymous referee for a first-class report. 
C.M.B. and A.L.  thank A.S.S.N.A. and the P.N.G. (France) for financial support. 
D.C.H. and S.D. warmly thank Observatoire de Strasbourg for hospitality during the course of this research. 

%                                                One column figure 
%----------------------------------------------------------- S_vib 

\begin{table*} 
\begin{center} 
\caption{Parameters of the models. All models have a total mass $M = 2\times 10^5 M_\odot$. 
 We set $R_h = R_g/2$ and $  \rho = \overline{\rho}$ in (\ref{eq:two-body}) to average $t_{ms}$.  
The subscript $0$ denotes $t=0$;  hats denote time-averages over the last 25 Myrs of evolution. 
 The last column gives the increase of concentration parameter $\delta c$ for a few cases (\S\S5.1, 5.2). }
\begin{tabular}{lccccccccc} 
 Name & Model & $t_{ms}$ &  $r_{ph,0}$ & $\widehat{r_{ph}}$  & \multicolumn{1}{c}{$\Sigma_0(0) $}    & \multicolumn{1}{c}{$\overline{\Sigma_0}(r_{ph}) $}    &  $\eta_o$  &  $\widehat{\eta} $  &   $\widehat{\delta c}$  \\ 
 & $\Psi/\sigma^2 $ & Myrs &   pc   & pc   &  $10^5 M_{\odot}/pc^{2}$   &  $10^5 M_{\odot}/pc^2$ &    &  &  \\  
\tableline\tableline \\ 
 `R136'   & $6$ &  9.3   & 1.01  &  0.51 $\pm$ 0.03  & 0.885  & 0.417 & 8.15      &  18.5 $\pm 1$ &  0.29  \\ 
 `M82-F'  &$6$ &  15.9 & 1.46   & 1.02  $\pm$ 0.06  & 0.474 & 0.224 & 8.15      &  12.6     $\pm 0.7$   &  \\ 
 \ Low     & $6$ &  47.9 &  3.05  & 2.84 $\pm$ 0.05   & 0.109 & 0.051 & 8.15      &    \   8.54    $\pm 0.06$   & \\ 
 \\
 `Plummer'  &$4.5$ & 7.5  & 1.05     &  0.62 $\pm$ 0.03 & 0.606 & 0.410  & 8.20   &  16 $\pm 1$ &  0.23 \\ 
                   &$9$       & 25.7 &  0.95    &  0.49 $\pm$ 0.02 &  5.232 & 0.440 & 7.76    &  19\ $\pm 4$ &  0.29 \\ 
\end{tabular} 
 \end{center} 
\end{table*} 

\newpage 

% ===================== Fig Eta vs time 

% ===================== Fig Eta vs time : END 

\end{document}